\documentclass[prd,floatfix,twocolumn,preprintnumbers]{revtex4}
\usepackage{graphicx}
\usepackage{amsmath,amssymb}
\input{epsf}
\usepackage{epsf}

\begin{document}

\def\be{\begin{equation}}
\def\ee{\end{equation}}

\title{Deviation From $\Lambda$CDM : Pressure Parametrization}
\author{A.A. Sen{\footnote{email:anjan.ctp@jmi.ac.in}}}
\affiliation{Center for Theoretical Physics, Jamia Millia Islamia, New Delhi 110025, India}
\date{\today}

\begin{abstract}
Most parametrizations for dark energy involve the equation of state $w$ of the dark energy.
 In this work, we choose the pressure of the dark energy to parametrize. As $p = constant$ 
 essentially gives a cosmological constant, we use the Taylor expansion around this behavior
  $p = -p_{0} + (1-a)p_{1} +...$ to study the small deviations from the cosmological constant.
   In our model, the departure from the cosmological constant behavior has been modeled by the
    presence of extra K-essence fields while keeping the cosmological constant term untouched.
     The model is similar to assisted inflation scenario in a sense that for any higher order
      deviation in terms of Taylor series expansion, one needs multiple K-essence fields. We 
      have also tested our model with the recent observational data coming from Supernova 
      type Ia measurements, the baryon oscillations peak (BAO) and the gas mass fraction of 
      the galaxy clusters inferred from X-ray observations and obtain constraints for
       our model parameters.     
\end{abstract}
\maketitle

\section{Introduction}
In recent years cosmology has become an interesting blend of an
empirical and theoretical sciences, combining concepts, tools and
observational results from astrophysics and particle physics. One
of the most important results in recent times is current
accelerated expansion of the universe rather than the
decelerating one as expected due to gravity. The clearest
evidence for this surprising result comes from the recent Type IA
Supernova  measurements \cite{super} that directly shows the acceleration. This apparent
acceleration of the universe  is attributed to a new, exotic energy density component with negative pressure which
also  balances the kinetic energy of the expansion. Recent observations of cosmic microwave background radiations (CMBR) by the WMAP experiment \cite{wmap} together with the data from the redshift surveys e.g Sloan Digital Sky Survey (SDSS) \cite{sdss} and the 2-Degree Field (2df) \cite{2df} redshift surveys for large scale structures have further strengthened this result.

Although a simple cosmological constant (C.C) or vacuum energy can serve the purpose for dark energy, it faces serious problem of fine tuning (for details, see the nice review by Padmanabhan \cite{paddy}). The other alternatives have been proposed, which includes an evolving scalar field called quintessence \cite{quint}, a scalar field with noncannonical kinetic term (K-essence) \cite{kess}, or simply a barotropic fluid with $p(\rho)$, such as the Chaplygin gas and its various generalizations \cite{gcg}. 

Despite having a large number of dark energy models without involving cosmological constant, the observational data still favor cosmological constant as a preferred dark energy candidate. But the interesting point is that it also allows a deviation from the corresponding C.C behavior. Although this deviation  is not large enough, it is still detectable by the present and future observational setups. The present work actually describe  a method  by which one can possibly model such small deviation from the cosmological constant behavior (also see a recent related work by Crittenden et.al \cite{critt} where similar thing has been discussed in terms of a minimally coupled scalar field).

\section{The Model Behavior}
    
We start with a fluid having a constant pressure $p = - p_{0}$ which ideally represents a positive cosmological constant. To investigate models which deviates slightly from this cosmological constant (C.C) behavior around present time, the simplest way is to do a Taylor series expansion around this behavior:

\be
p = -p_{0} + (1-a) p_{1}
\label{first}
\ee
where we have assumed the present day scale factor $a=1$ without any loss of generality and $p_{1} = {dp\over{da}}(a=1)$. We have taken the minus sign infront the $p_{0}$ term as at present ($a=1$), the dark energy pressure should be negative (here we have assumed $p_{0} > 0$). This is simply a choice of notation and does not affect any of our conclusions. In the above and subsequent equation we have not assumed anything about the sign of $p_{1}$. We shall show later that some combinations of $p_{0}$ and $p_{1}$ can be observationally constrained instead of $p_{0}$ and $p_{1}$ individually. We have also considered term up to the first order in the series expansion. This is similar to the widely used parametrization for the equation of state (e.o.s) for the dark energy $w = -w_{0} + (1-a)w_{1}$ first introduced by Chevallier and Polarski \cite{pol} and extensively studied by Linder \cite{linder1}. The only difference is that the zeroth order behavior in (1) is essentially a C.C whereas for the above e.o.s parametrization, the zeroth order is C.C only when one assumes $w_{0} = -1$. The importance of parametrizing dark energy using its pressure has also been discussed by Sahni et.al \cite{sahni1} in terms of the statefinder parameter $s$ which is extremely sensitive to the pressure (also see \cite{sahni2} for a recent review on dark energy reconstruction).  One can also write the equation (\ref{first}) as

\be
p = -(p_{0} - p_{1}) - p_{1} a.
\label{first2}
\ee

Now one can use the energy conservation equation ${d\rho\over{dt}} + 3H(\rho+p) =  0 $, together with equation (\ref{first}), to find the energy density for the dark energy fluid as

\be
\rho = (p_{0} - p_{1}) + (3/4) p_{1} a,
\label{denfirst}
\ee
where we have set the integration constant to zero. keeping the integration constant non-zero, would result a term which scales as $\sim a^{-3}$. But as we already assume the presence of matter-like component separately in our model, we prefer to ignore the integration constant. This does not change any of our conclusions. Equation (\ref{denfirst}) together with equation (\ref{first}) represents a C.C together with another fluid with equation of state $w = -4/3$ which is necessarily a phantom one. This shows that when one parametrizes C.C through its pressure, the first order deviation necessarily gives rise to an extra term with phantom equation of state. This result is quite general without any added assumption. This may be due to the pressure parametrization of the small deviations from the C.C behavior.

Going to the second order for the Taylor series expansion, one can write the pressure term as

\be
p = -p_{0} + (1-a)p_{1} + {1\over{2}}(1-a)^{2} p_{2},
\label{second}
\ee
where $p_{2} = {d^2p\over{da^2}}(a=1)$. This expression can be rewritten as

\be
p = -(p_{0}-p_{1}-p_{2}) - (p_{1}+p_{2})a + {1\over{2}}p_{2}a^2.
\label{second1}
\ee

Using equation (\ref{second}), one can now integrate the energy conservation equation to get the expression for the energy density for the dark energy:

\be
\rho = (p_{0} - p_{1} - p_{2}/2) +(3/4)(p_{1}+p_{2})a - (3/10)p_{2}a^2.
\label{densec}
\ee

Assuming $p_{2} < 0$, to satisfy the positivity condition for the dark energy density at all times, equation (\ref{second1}) and equation (\ref{densec}) now represents a C.C plus two more fluid behavior with equation of states  -4/3 and -5/3. In general it is very easy to check that for each extra order $n$ in the Taylor series expansion around the C.C behavior ($p = -p_{0}$ = constant), there is an extra fluid component with equation of state $ w = [-1-{n\over{3}}]$. The important thing is that all these extra fluid behaviors are necessarily phantom type.

To model such behavior, standard minimally coupled scalar field with canonical kinetic energy is not suitable as one can  not achieve phantom type equation of state with this type of scalar field.  The other simple possibility is to consider the K-essence type scalar fields \cite{kess} with non-canonical kinetic energy term (there can be other alternatives like non-minimally coupled scalar fields or $f(R)$ theories, which we are not considering here). This has been widely considered for modeling dark energy especially with phantom type equation of state. Using such K-essence type scalar field, one can show that each extra order deviation in the Taylor series expansion around the C.C behavior can be modeled with a K-essence field with Lagrangian

\be
{\cal{L}} = - cX^{n/(2(3+n))}
\label{lagran}
\ee
where $X = g^{\mu\nu}\phi_{,\mu}\phi_{,\nu}$ is the kinetic energy of the field (here $(,\mu)$ means derivative w.r.t $x^{\mu}$) and $n$ is the order of expansion and $c$ is a constant. So in general if one makes a Taylor series expansion around the C.C behavior as

\be
p = -\Lambda + (1-a) p_{1} + (1/2)(1-a)^2 p_{2} + ......
\label{expan}
\ee
then the corresponding the Lagrangian can be written as

\be
{\cal{L}} = -\Lambda - \sum_{i=1}^{n} c_{i} X_{i}^{i/2(3+i)}.
\label{tlag}
\ee
Here $X_{i}$ is the kinetic energy for i-th field $\phi^{i}$ needed for the i-th order expansion, $X_{i} = g^{\mu\nu}\phi^{i}_{,\mu}\phi^{i}_{,\nu}$ and $c_{i}$ is the constant associated with the i-th field. This means for n-th order expansion in the Taylor series, one needs $n$ number of fields.
This is similar to the Assisted inflation scenario \cite{assisted}, where one needs multiple fields for sufficient e-folding for inflation. Here also the more one deviates from the C.C behavior towards super acceleration regime (due to phantom nature of the extra fluid behavior), one needs to include more k-essence type scalar fields in the total Lagrangian. It can be termed as ``Assisted super acceleration''.
But current observational data do not predict a large deviation from the C.C behavior for dark energy, and only allows a marginal departure from the C.C. behavior. Hence it is sufficient to consider only up to the first order in the Taylor series expansion given by equation (\ref{first}). In that case one needs only one K-essence field  and the Lagrangian for this purpose can be written as

\be
{\cal{L}} = -\Lambda - c_{1} X_{1}^{1/8}.
\label{lag1}
\ee

\begin{figure}[t]
\centerline{\epsfxsize=3.2truein\epsffile{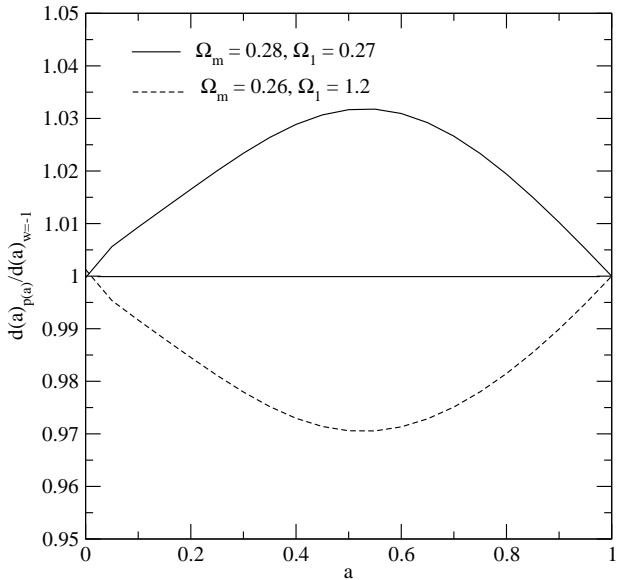}}
\caption{Convergence of the distance-redshift relation by matching the distance to CMB last scattering for models with different values of $\Omega_{m}$ and corresponding $\Omega_{1}$ to the $\Lambda$CDM case.}
\end{figure}
One can also think of equation (\ref{first2}) and (\ref{denfirst}), as a single dark energy fluid with equation of state

\be
w = -{{(p_{0}-p_{1})+p_{1}}a\over{(p_{0}-p_{1})+(3/4)p_{1}a}}
\label{eos}
\ee
which is similar to Pade approximation of order (1,1) \cite{pade}.It is also known that Pade approximation often gives better approximation of the function than truncating its Taylor series and it may still work where the Taylor series does not converge.   Assuming $p_{1} > 0$ the equation of state of this  single dark energy fluid is also phantom in nature as for any $a>0$, the numerator of the expression in the r.h.s of (\ref{eos}) is always greater than the denominator. 

Next we write the expression for the Hubble parameter, by considering only up to first order for the Taylor series expansion for the pressure term around the C.C behavior (\ref{first}):

\be
H(z)^2 = H_{0}^2 [\Omega_{m}a^{-3} + \Omega_{1} + \Omega_{2}a]
\label{hubble}
\ee
where $H_{0}$ is the present day value for the Hubble parameter, $\Omega_{m}$ is the density parameter for the matter energy density, $\Omega_{1} = (p_{0}-p_{1})/(3H_{0}^2/8\pi G)$ and $\Omega_{2} = (3/4)p_{2}/(3H_{0}^2/8\pi G)$. Here we consider only the flat universe, hence $\Omega_{m}+\Omega_{1}+\Omega_{2} = 1$. So we have only two free parameters in our model. Recently Linder \cite{linder} has shown that high redshift distance measurements (like distance to last scattering for CMB) consistent with LCDM, virtually forces the low redshift measurements like SNIa two measure $w =-1$ for dark energy despite the presence of varying $w(z)$. For this purpose, Linder has used the equation of state parametrization $w = w_{0} + (1-a)w_{a}$ for the dark energy. Here we check this for our dark energy parametrization given by equation (\ref{first}) where we use pressure of the dark energy instead of its equation of state. The result has been shown in Figure 1 which confirms the Linder's result in \cite{linder}. It also shows that the result is independent of dark energy parametrization. By using a different parameter e.g the pressure, we reach the same conclusion. 

\begin{figure}[t]
\centerline{\epsfxsize=3.5truein\epsffile{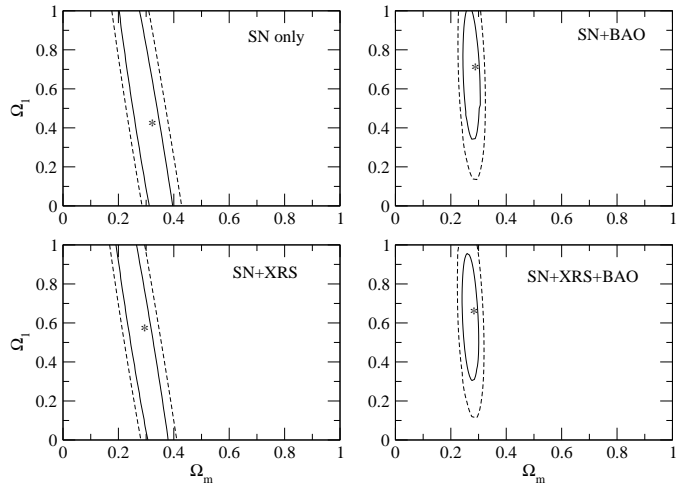}}
\caption{$1\sigma$ (solid line) and $2\sigma$ (dashed line)confidence for various observational data discussed in paper. The '*'' is the best fit point for each data fitting also described in the paper.}
\end{figure}

\section{Fitting With the Observation}

Our goal is to constrain our model (\ref{hubble}), using different cosmological data. For this purpose we use the data from the various SNIa observations in recent times. In particular we use 60 Essence supernovae \cite{super},
57 SNLS (Supernova Legacy Survey) and  45 nearby supernovae.
We have also included the new data release of 30 SNe Ia detected by
HST and classified as the Gold sample by Riess et al \cite{super}.
The combined data set can be found in Ref. \cite{davis}. The total number of data points involved is 192.  The best fit values for our parameters $\Omega_{m}$ and $\Omega_{1}$ are given by $\Omega_{m} = 0.31$ and $\Omega_{1} = 0.38$ with $\chi^{2}_{min}/(d.o.f) = 1.03$. There is a large contribution (around $31\%$) from the extra phantom component. 

Next we add the measurement of the CMB (Cosmic Microwave Background) acoustic scale  at $z_{BAO} = 0.35$ as observed by the SDSS (Sloan Digital Sky Survey) for the large scale structure ( This is the Baryon Acoustic Oscillation (BAO) peak) \cite{sdss}. Considering this data together with the SNIa data discussed in the previous paragraph, the best fit values for the model parameters become $\Omega_{m} = 0.27$ and $\Omega_{1} = 0.67$ with the $\chi^{2}_{min}/(d.o.f) = 1.024$ which is a slight improvement from the SNIa only data. Also adding the BAO data makes the model more close to the $\Lambda$CDM than considering only SNIa data as the contribution from extra phantom term in equation (\ref{hubble}) becomes much less (around $5\%$). 

We also consider the gas mass fraction of galaxy cluster , $f_{gas} = M_{gas}/M_{tot}$ inferred from the X-ray observations \cite{xrs}. This depends on the angular diameter distance $d_{A}$ to the cluster as $f_{gas} = d_{A}^{3/2}$. The number of data point involved is 26. When we add these data with SNIa data discussed previously, the best fit values for the model parameters become $\Omega_{m} = 0.28$ and $\Omega_{1} = 0.53$ with $\chi^{2}_{min}/(d.o.f) = 1.013$ which is also a marginal improvement from the previous two cases. But there is also sufficient contribution (around $19\%$) from the extra phantom part. 

When we add all the data mentioned above (SNIa + BAO + XRS), the best fit values for the parameters become $\Omega_{m} = 0.27$ and $\Omega_{1} = 0.62$ with the $\chi^{2}_{min}/(d.o.f) = 1.008$ which is sufficient improvemnet from the all the above cases. the contribution from the extra phantom component is also not much (around $11\%$). This together with the other results discussed above, shows that including the BAO data put tight constraint on the extra phantom part in our model. In other words, BAO data severely restricts models that deviates significantly from the $\Lambda$CDM model. In Figure 1, we have shows the $1\sigma$ and $2\sigma$ confidence contours for our model parameters $\Omega_{m}$ and $\Omega_{1}$ for different observational data discussed above.

\section{Conclusion}

In conclusion we have studied  dark energy models which are very close the $\Lambda$CDM behavior. In doing so, we parametrize the dark energy through its pressure $p$  instead of its equation of state $w$. As $p = constant$ ideally represents the C.C, in order to study models which are very close to this C.C behavior around present time, we have used the most simple procedure of Taylor expanding $p$ around its C.C behavior. This gives series of terms in the dark energy density expression, with phantom type equation of states. The interesting point is that one can find a relation for the equation of state (e.o.s) of these individual term given by $w_{n} = -1 - n/3$. As all these extra terms are phantom in nature, we have modeled this extra fluids by K-essence fields and have shown that as one goes for higher order deviations from the C.C behavior, one needs more and more extra K-essence fields. This is similar to ``Assisted Inflation'' scenario where one needs more than one  inflaton field to obtain higher e-folding which may not be obtained using a single field. Our scenario is like ``Assisted Super Acceleration'' where in order to super accelerate more ( i.e deviating more towards phantom region), one needs more and more extra K-essence fields. The combined Lagrangian for the dark energy part which deviates up to n-th order from the C.C behavior has been given in equation (9).

Although we have used different parametrization for the dark energy, we have confirmed the recent conclusion by Linder \cite{linder} that there is always a convergence in the distance redshift relation at low redshift, if one matches the distance to CMB last scattering for any dark energy model to the $\Lambda$CDM model.

We have also constrained the parameters $\Omega_{m}$ and $\Omega_{1}$ by fitting our model with different observational data e.g supernova data, BAO data and also the data coming from the gas mass fraction of galaxy clusters from the X-ray observation. For this, we have assumed only the first order variation from the C.C behavior. The result shows that adding BAO data severely constrains the extra term and make the model very close to $\Lambda$CDM.

Recent advances in observational cosmology have put tight constraint on the properties of dark energy. $\Lambda$CDM is emerging out to be more favorable model for various observational data although one has to take extra precautions before making any such conclusion (See recent work by \cite{linder} for this purpose). But the current observational data also allow us to consider models that are very close to $\Lambda$CDM but still different from that (the most recent bound on the effective e.o.s for dark energy is $w = -1 \pm 0.1$ (see \cite{linder} and references therein). One way is to consider such models is to assume that C.C is exactly zero and consider time dependent dark energy behaviors. This has been the standard practice for most of the investigations concerning dark energy model building. The other possibility is to assume the existence of C.C with small but non-zero value and consider extra component to model the observed departure from C.C.  This is because of the fact that we do not have any known mechanism to make the C.C vanish, without direct conflict with meaningful physics \cite{cc}. Also recent advances in string-landscape models have resulted in explaining the extremely small but non-zero value of C.C that is required by the observations (See \cite{bousso} and references therein for details). Keeping this in mind, we consider the second possibility, which may be first approach of this kind. We think further investigations pursuing this kind of approach can yield interesting consequences in dark energy model building.

\vspace{10mm}
\section{Acknowledgement}
The author is grateful to Harish-Chandra Research Institute, Allahabad, India and Abdus Salam International Center for Theoretical Physics, Trieste, Italy for their hospitality where part of this work has been carried out. The author is also grateful to J.S. Bagla for his suggestions and comments during the initial stages of this work. The author also acknowledges the useful discussions with J. Majumdar.

\end{document}